\documentstyle[12pt]{article}
\begin{document}
\title{ Density Effect on Hadronization of a Quark Plasma}
\author{P. Zhuang, M. Huang and Z. Yang\\
        Physics Department, Tsinghua University, Beijing 100084, China}
\date{}
\maketitle

\begin{abstract}
The hadronization cross section in a quark plasma at finite temperature and density is calculated in the framework of Nambu--Jona-lasinio model with explicit chiral symmetry breaking. In apposition to the familiar temperature effect, the quark plasma at high density begins to hadronize suddenly. It leads to a sudden and strong increase of final state pions in relativistic heavy ion collisions which may be considered as a clear signature of chiral symmetry restoration. 
\end{abstract}

\section { Introduction }

It is generally believed that there are two QCD phase transitions in hot and dense nuclear matter\cite{MuHe}. One of them is related to the deconfinement process in moving from a hadron gas to a quark-gluon plasma, and the other one describes the transition from the chiral symmetry breaking phase to the phase in which it is restored. The theoretical evidence from the lattice simulations\cite{Karsch} of QCD at finite temperature for these phase transitions has prompted experimental efforts to create the new phases of matter in the laboratory during the early stages of relativistic heavy ion collisions. Since it is still difficult to extract definite information about  the phase transitions from the lattice simulations with nonzero baryon density, we need QCD models to investigate the phase transitions at finite density. From the recent study \cite{BeRa} based on the instanton model, the phase structure is more rich at high density than at high temperature. There exists a new phase of color superconductivity when the density is sufficiently high. 

Chiral symmetry plays a crucial role for the behavior of light hadrons, in determining not only their static properties in the vacuum, but also the changes therein that are due to in-medium effects. One of the models that enables us to see directly how the dynamic mechanisms of chiral symmetry breaking and restoration operate is the Nambu--Jona-Lasinio (NJL) model applied to quarks\cite{NJL}. Within this model, one can obtain the hadronic mass spectrum and the static properties of mesons remarkably well. In particular, one can recover the Goldstone mode, and some important low-energy properties of current algebra such as the Goldstone-Treiman and Gellmann-Oakes-Renner relations. Because of the contact interaction between quarks that is introduced in the model, there is of course no confinement. A further consequence of this feature is that the model is non-renormalizable, and it is necessary to introduce a regulator $\Lambda$ that serves as a length scale in the problem, and which can be thought of as indicating the onset of asymptotic freedom. 

At low baryon density, the phase transition of chiral symmetry restoration in chiral limit is of second order, in the case with quarks of nonzero mass it becomes a smooth crossover as a function of temperature and density. This picture obtained in the NJL model with two flavors is consistent with present lattice simulations and many models\cite{WiRa}. However, at high density several models\cite{BeRa,NJL} suggest that the chiral symmetry restoration transition is of first order. Assuming that this is the case in QCD, this first order transition will remain in the case in which explicit chiral symmetry breaking is present. This raises the possibility that some consequences of chiral symmetry restoration obtained in the chiral limit may be found in the real world with massive pions. Some authors\cite{Hei} have discussed the signatures of a first-order phase transition, such as the event-by-event rapidity fluctuations and the associated variations in HBT radii. 

In relativistic heavy ion collisions, nuclear matter is first heated and/or compressed up to a region of both a chirally symmetric and a deconfined phase of matter, and then in the cooling process the deconfined partons will recombine into the observed hadrons, which are mainly pions. It is important to understand how a quark plasma converts into a hadron gas at finite temperature and density, and especially what the influence of the chiral symmetry restoration transition is on the hadronization process. Although we are aware of the disadvantages of the NJL model, the lack of confinement and non-renormalizability of the effective 4-fermion interaction, its strengths, the transparent description of the chiral phase transition and the binding of quarks into mesons, it is worthwhile to discuss the effect of chiral phase transition on the hadronization rate in this model. From the study\cite{Hue} of the hadronization rate for the conversion of a quark-antiquark pair into two pions, $q\bar q\rightarrow 2\pi$ at zero baryon density and in the chiral limit, the hadronization cross section diverges at the critical temperature $T_c$ of the chiral phase transition. Since the phase transition at low density will be fully washed out when  explicit chiral symmetry breaking is taken into account, the dynamic divergence will no longer exist in the real world. A natural question is whether any information about the chiral phase transition at high density is retained in the hadronization rate when a nonzero current quark mass is considered. In this paper, we investigate at finite temperature and density the hadronization rate for $q\bar q\rightarrow 2\pi$ in the flavor $SU(2)$ NJL model with explicit chiral symmetry breaking.

The paper is organized as follows. In Section 2, we briefly review the quarks at the mean field level and mesons within the  random phase approximation (RPA). We  discuss especially the phase diagrams of the model in temperature-chemical potential ($T-\mu$) plane and temperature-baryon density ($T-n_b$) plane. In Section 3, we study the behavior of the hadronization cross section to first order in a $1/N_c$ expansion as a function of the c.m. energy $s$, temperature $T$ and baryon density $n_b$. We focus our attention on the difference between temperature and density effects, and the influence of the chiral phase transition on the hadronization rate. We summarize in Section 4. 

\section {Phase Diagrams }

The two-flavor version of the NJL model is defined through the Lagrangian density,
\begin{equation}
\label{njl}
L_{NJL} = \bar\psi (i\gamma^\mu\partial_\mu - m_0)\psi+G[(\bar\psi\psi)^2+(\bar\psi i \gamma_5\tau\psi)^2],
\end{equation}
where only the scalar and pseudoscalar interactions corresponding to $\sigma$ and $\pi$ mesons, respectively, are considered, $\psi$ and $\bar\psi$ are the quark fields, $\tau$ is the $SU(2)$ isospin generator, $G$ is the coupling constant with dimension $GeV^{-2}$, and $m_0$ is the current quark mass.

The key quantity for dealing with the thermodynamics of a variable number of particles is the thermodynamic potential $\Omega$. In the mean field approximation, which is the lowest order $O(1)$ in the $1/N_c$ expansion, it is\cite{Zh}
\begin{eqnarray}
\label{omega}
\Omega(T,\mu, m_q) && ={(m_q-m_0)^2\over 4 G}-2N_c N_f\int{d^3{\bf p}\over (2\pi)^3} E_p\nonumber\\
&& -2 N_c N_f T \int {d^3{\bf p}\over (2\pi)^3}\ln\left(1+e^{-{E_p+\mu\over T}}\right)\left(1+e^{-{E_p-\mu\over T}}\right)\ ,
\end{eqnarray}
where $E_p^2={\bf p}^2+m_q^2$, $\mu$ is the chemical potential, and $N_c, N_f$ are the number of colors and flavors respectively. Equation (\ref{omega}) holds for any value of the quark mass $m_q$, which is the order parameter of the chiral phase transition. In the spirit of thermodynamics, the physical system is described by that value of $m_q$ for which $\Omega(T,\mu, m_q)$ is a minimum, 
\begin{equation}
\label{mini}
{\partial\Omega\over \partial m_q}=0\ ,\ \ \ \ \ \ \ {\partial^2\Omega\over\partial m_q^2}\ge 0\ .
\end{equation}
The above equality is the socalled gap equation to determine the constituent quark mass as a function of $T$ and $\mu$. The inequality is especially necessary in high density region where the thermodynamic potential has two minima and one maximum. 

In the chiral limit when the current quark mass $m_0 = 0$, the solution of the gap equation separates the region of chiral symmetry breaking with $m_q \ne 0$ from the region of chiral symmetry restoration with $m_q = 0$. The phase-transition lines demarcating these two regions are given in Fig.1a in the $T-\mu$ plane and Fig.1b in the $T-n_b$ plane. The relation between the baryon density $n_b$ and the chemical potential $\mu$ is rather complicated and is also a function of $T$,
\begin{equation}
\label{nb}
n_b = {N_c N_f\over 3}\int{d^3{\bf p}\over (2\pi)^3}\left[\tanh{1\over 2T}(E_p +\mu)-\tanh{1\over 2T}(E_p-\mu)\right]\ ,
\end{equation}
where the factor $3$ reflects the fact that three quarks make a baryon.

The order of the chiral phase transition is determined by the behavior of the order parameter across the phase-transition line. If the quark mass $m_q$ is continuous, the phase transition is of second order indicated by dashed lines in Fig.1, otherwise it is of first order shown by thin solid lines. The tricritical point $P$ which separates the first- and second-order phase transitions is located at $T=0.079\ GeV$ and $\mu = 0.28\ GeV$ or $n_b/n_0 = 1.95\ $ with $n_0 = 0.17/fm^3$ being the baryon density in normal nuclear matter. Its positions in the $T-\mu$ and $T-n_b$ planes are close to that calculated in the Landau $\phi^6$ model and the instanton model\cite{Ra}.

At any critical point on the first-order phase-transition line in the $T-\mu$ plane, the thermodynamic potential as a function of the quark mass has two minima and one maximum between them. The coexistence of two minima indicates a mixed phase of chiral symmetry breaking and chiral symmetry restoration. The two spinodal points on the left hand and right hand sides of the maximum are characterized by a stability equal to zero,
\begin{equation}
\label{spin}
{\partial^2\Omega\over \partial m_q^2}=0\ .
\end{equation}
Between the spinodal points, ${\partial^2\Omega\over \partial m_q^2}<0$, the system is unstable with respect to a small fluctuation in quark mass. Outside the spinodal points and inside the minima, ${\partial^2\Omega\over \partial m_q^2}>0$, the system is metastable. In these regions a finite fluctuation in $m_q$ may be amplified, and the system may not return to its original state after a sufficient deviation. Outside of the minima, ${\partial^2\Omega\over \partial m_q^2}>0$, the system is stable. The unstable and metastable regions are marked by $US$ and $MS$ in Fig.1b.

In the real world with nonzero current quark mass, the second-order phase transition becomes a smooth crossover and the tricritical point $P$ becomes a critical end-point $E$ of a first-order phase-transition line. The end-point $E$ is a second-order critical point and shifted with respect to the tricritical point $P$ towards larger $\mu$ in the $T-\mu$ plane and towards smaller $n_b$ in the $T-n_b$ plane. The first-order phase-transition lines are denoted by thick solid lines in the phase diagrams. Their shift with respect to the case of chiral limit is due to the fact that it is more difficult for a phase transition to happen in a system of massive particles than that in the corresponding system of massless particles. 

In the calculations of the phase-transition lines shown in Fig.1, the $3$ parameters in the NJL model, namely the coupling constant $G$, the momentum cutoff $\Lambda$ and the current quark mass $m_0$, are fixed by the constituent quark mass $m_q =0.32\ GeV$, the pion decay constant $f_\pi = 0.093\ GeV$ and the pion mass $m_\pi = 0.134\ GeV$ (in the chiral limit $m_\pi = 0$) in the vacuum ($T=\mu = 0$). In the self-consistent mean field approximation, it is well known that the masses of the $\pi$ and $\sigma$ mesons satisfy the dispersion relation\cite{NJL}
\begin{equation}
\label{dispersion}
1-2G\Pi_M (T,\mu,m_M) = 0\ ,
\end{equation}
where $m_M$ is the mass of the respective meson and $\Pi_M$ is the associated retarded polarization function,
\begin{eqnarray}
\label{polar}
&& \Pi_\pi (T,\mu,m_\pi) = N_c N_f \int{d^3{\bf p}\over (2\pi)^3} {1\over E_p}{E_p^2\over E_p^2-m_\pi^2/4}\left[\tanh{1\over 2T}(E_p+\mu)+\tanh{1\over 2T}(E_p-\mu)\right]\ ,\nonumber\\
&& \Pi_\sigma (T,\mu,m_\sigma) = N_c N_f \int{d^3{\bf p}\over (2\pi)^3} {1\over E_p}{E_p^2-m_q^2\over E_p^2-m_\sigma^2/4}\left[\tanh{1\over 2T}(E_p+\mu)+\tanh{1\over 2T}(E_p-\mu)\right]\ .
\end{eqnarray}
The pion coupling constant $g_\pi$ is related to the residue at the pole of the pion propagator,
\begin{eqnarray}
\label{gpi}
g_\pi^{-2} && ={\partial \Pi_\pi(T,\mu,m_\pi)\over \partial m_\pi^2}\nonumber\\
&& ={N_c N_f\over 4}\int{d^3{\bf p}\over (2\pi)^3}{E_p\over (E_p^2-m_\pi^2/4)^2}\left[\tanh{1\over 2T}(E_p+\mu)+\tanh{1\over 2T}(E_p-\mu)\right]\ .
\end{eqnarray}
It is easy to see that when $m_\pi > 2 m_q,\ g_\pi = 0$, the pions will decay into quark-antiquark pairs, $\pi\rightarrow q\bar q$. Therefore, we can determine a hadronization line by 
\begin{equation}
\label{hadro}
m_\pi(T_h,\mu_h) = 2 m_q(T_h,\mu_h)\ .
\end{equation}
For temperature $T\ge T_h$ and chemical potential $\mu \ge \mu_h$ or baryon density $n_b\ge n_{bh}$, mesons become resonant states and no more pions exist in the model, and one has a system of interacting quarks. For $T\le T_h$ and $\mu\le \mu_h$ or $n_b\le n_{bh}$, a quark-meson plasma is formed. This means that in the cooling process of relativistic heavy ion collisions the quark plasma begins to hadronize at the point $(T_h, n_{bh})$. The coexistence of quarks and mesons at low temperature and density is of course an artifact of the NJL model.  
 
In the chiral limit, the equation (\ref{hadro}) for determining the hadronization line is fully equivalent to the equation for defining the chiral phase-transition line. Therefore, the two lines coincide in the phase diagrams. When explicit chiral symmetry breaking is taken into account, the two equations become different. While there is no longer a phase transition at high temperature, one can still define the hadronization lines via equations (\ref{hadro}) and (\ref{nb}) which are shown in Fig.1.   
\section { Hadronization Cross Section }

The total hadronization cross section of a single quark in a flavor $SU(2)$ model is
defined as\cite{Hue}
\begin{equation}
\label{cross1}
\sigma_q = \sigma_{u\bar u\rightarrow 2\pi_0}+\sigma_{u\bar u\rightarrow \pi_+\pi_-}\sigma_{u\bar d\rightarrow \pi_+\pi_0}\ ,
\end{equation}
this is a measure of how quickly a quark hadronizes from a charge symmetric plasma. Since the NJL model is a strong-coupling theory, perturbation theory is inapplicable, and we require a selection procedure for the relevant Feynman graphs. We choose the $1/N_c$ expansion, and work to the first order. To $O(1/n_c)$, the relevant T-matrix amplitudes for the reaction $q\bar q\rightarrow 2\pi$ are sketched in Fig.2. It depicts the s-channel amplitude $T^s$ in which an intermediate $\sigma$-meson is produced, and the t-channel which displays a quark exchange. It is understood that the crossed diagrams are included for a process in which the final state particles are identical, e.g. for $u\bar u\rightarrow 2\pi_0$.

Using an obvious notation for the T-matrix amplitudes, one has for the process $u\bar u\rightarrow 2\pi_0$ the s-channel amplitude
\begin{equation}
\label{schannel}
T^s = \bar v(p_2)u(p_1){2G\over 1-2G\Pi_\sigma (p_1+p_2)}g_\pi^2 A_{\sigma\pi\pi}(p_1+p_2,p_3)\delta_{c_1 c_2}\ ,
\end{equation}
where $u$ and $v$ are quark and antiquark spinors, the symbol $\delta_{c_1 c_2}$ refers to the color degree of freedom, and $A_{\sigma\pi\pi}$ represents the amplitude of the triangle vertex $\sigma\rightarrow\pi\pi$. By assuming the $q\bar q$ system to be at rest in the heat bath and evaluating the Matsubara sum in the triangle, one obtains the explicit dependence of the amplitude $A_{\sigma\pi\pi}$ on the Mandelstam Lorentz scalar $s=(p_1 + p_2)^2$, and the temperature $T$ and chemical potential $\mu$,
\begin{eqnarray}
\label{trinagle}
A_{\sigma\pi\pi}(s,T,\mu) =&&  4m_q N_c N_f \int{d^3{\bf p}\over (2\pi)^3}{f_F(E_p-\mu)-f_F(-E_p-\mu)\over 2E_p}\times \nonumber\\
&& {8({\bf p}\cdot{\bf p}_3)^2-(2s+4m_\pi^2){\bf p}\cdot{\bf p}_3+s^2/2-2sE_p^2\over (s-4E_p^2)\left((m_\pi^2-2{\bf p}\cdot{\bf p}_3)^2-E_p^2 s\right)}\ ,
\end{eqnarray}
where ${\bf p}_3^2 = \sqrt{s/4-m_\pi^2}$, and the Fermi-Dirac distribution function $f_F(x) = \left(1+e^{x/T}\right)^{-1}$. In terms of the other two Mandelstam variables $t=(p_1-p_3)^2$ and $u=(p_1-p_4)^2$ with the constraint $s+t+u=2m_q^2+2m_\pi^2$, the direct and exchange contributions to the t-channel amplitude are
\begin{eqnarray}
\label{tuchannel}
&& T^{t,dir} = \bar v(p_2)\gamma_5{\gamma^\mu (p_1-p_3)_\mu +m_q\over t-m_q^2}\gamma_5u(p_1)g_\pi^2\delta_{c_1 c_2}\ ,\nonumber\\
&& T^{t,exc} = \bar v(p_2)\gamma_5{\gamma^\mu (p_1-p_4)_\mu +m_q\over u-m_q^2}\gamma_5u(p_1)g_\pi^2\delta_{c_1 c_2}\ .
\end{eqnarray}
Since the s-channel crossed graph equals the direct term, the differential cross section can be written as
\begin{equation}
\label{cross2}
{d\sigma_{u\bar u\rightarrow 2 \pi_0}\over dt}(s,t,T,\mu) = {1\over 16\pi s(s-4m_q^2)}\sum_{c,s}'|2T^s (s,T,\mu)+T^{t,dir}(t,T,\mu)+T^{t,exc}(u,T,\mu)|^2\ ,
\end{equation}
where the prime on the summation indicates the average over spin and color degrees of freedom of the quarks in the initial state.

It is necessary to note that in the s-channel the triangle amplitude $A_{\sigma\pi\pi}$ and the $\sigma$ polarization function $\Pi_\sigma$ are both complex functions. Their real and imaginary parts are separated from each other by using the decomposition ${1\over x+i\epsilon} = P {1\over x}+i\pi\delta(x)$ for $A_{\sigma\pi\pi}(s+i\epsilon, T,\mu)$ and $\Pi_\sigma(s+i\epsilon,T,\mu)$.

After performing the integration over the momentum transfer $t$, the integrated cross section is obtained from
\begin{equation}
\label{cross3}
\sigma_{u\bar u\rightarrow 2\pi_0}(s,T,\mu) = \int_{t_{min}}^{t_{max}}dt {d\sigma_{u\bar u\rightarrow2\pi_0}\over dt}(s,t,T,\mu)\left(1+f_B({\sqrt s\over 2})\right)^2\ ,
\end{equation}
where we have the limits $t_{max} = -{1\over 4}(s-4m_q^2)-{1\over 4}(s-4m_\pi^2)+{1\over 2}\sqrt{(s-4m_q^2)(4-m_\pi^2)}$, and $t_{min} = -{1\over 4}(s-4m_q^2)-{1\over 4}(s-4m_\pi^2)$ for the process $u\bar u\rightarrow 2\pi_0$ and $t_{min} = -{1\over 4}(s-4m_q^2)-{1\over 4}(s-4m_\pi^2)-{1\over 2}\sqrt{(s-4m_q^2)(4-m_\pi^2)}$ for the other two processes. In equation (\ref{cross3}) we have taken into account the Bose-Einstein distribution function $f_B(x)=\left(e^{x/T}-1\right)^{-1}$ for the final state pions. 

In the same way as equation (\ref{cross2}) was constructed, we can write the differential cross sections for the processes $u\bar u\rightarrow \pi_+\pi_-$ and $u\bar d\rightarrow \pi_+\pi_0$ explicitly as
\begin{eqnarray}
\label{cross4}
&& {d\sigma_{u\bar u\rightarrow \pi_+\pi_-}\over dt}(s,t,T,\mu) = {1\over 16\pi s(s-4m_q^2)}\sum_{c,s}'|T^s (s,T,\mu)+2T^{t,dir}(t,T,\mu)|^2\ ,\nonumber\\
&& {d\sigma_{u\bar d\rightarrow \pi_+\pi_0}\over dt}(s,t,T,\mu) = {1\over 16\pi s(s-4m_q^2)}\sum_{c,s}'|\sqrt 2 T^{t,dir}(t,T,\mu)|^2\ .
\end{eqnarray}

The total hadronization cross section discussed above in the whole $T-\mu$ (or $T-n_b$) plane and with explicit chiral symmetry breaking was studied in the chiral limit and along the $T$ axis in Ref.\cite{Hue}. The main result is the singularity structure of the cross section: $\sigma_q(s,T)$ displays singularities at the threshold energy $s_{th}=4m_q^2$ for any $T$ and at the critical temperature $T_c$ for any $s$. The singularity at $s_{th}$ results from the fact that the entrance channel has a threshold energy $s_{th}$, while the exit channel has $s=0$. The singularity at $T_c$ arises from the chiral symmetry restoration reflected on the quark exchange in the t-channel: As $T\rightarrow T_c,\ m_q\rightarrow 0$, and then the amplitude $T^{t,dir}\rightarrow\infty$. 

What we want to stress in this paper is the baryon density effect and the $m_\pi$ dependence of the hadronization rate. Fig.3 shows the energy dependence of the integrated total cross section as a function of energy $s$ for three values of the density $n_b$ at zero temperature (Fig.3a) and for three values of the temperature $T$ at zero density (Fig.3b). Because of the three-momentum cutoff $\Lambda$, the energy range is restricted to $4 \left(Max(m_q,m_\pi)\right)^2\le s \le 4\left(\left(Max(m_q,m_\pi)\right)^2+\Lambda^2\right)$. At low densities, there is still a kinematical divergence at $s_{th} = 4 m_q^2$ when $m_q>m_\pi$. For ${n_b\over n_0}=1.1$ shown by thin solid line in Fig.3a, the cross section drops down from the infinity at 
$s_{th}=1.33\ GeV^2$ and then goes up again and reaches the second infinity at $s=m_\sigma^2$ which is the contribution from the $\sigma$-exchange in the s-channel. Then the cross section decreases and is fairly constant for large energies. With the increase of the density, when $m_\pi > m_q$, the threshold energy of the reaction is larger than the entrance energy, and therefore the first singularity disappears. In Fig.3a this is the case for ${n_b\over n_0}=1.25$ denoted by dashed line. Further when the density is sufficiently high that the threshold energy $s=4m_\pi^2$ of the reaction is larger than $m_\sigma^2$, the second singularity disappears too, the cross section has no further singularities, see the thick solid line for ${n_b\over n_0}=1.5$ in Fig.3a. The energy dependence of the cross section at finite temperature is similar and shown in Fig.3b.

The energy dependence of the singularity structure discussed here is very different from that in the chiral limit where there is always only one singularity at the threshold $s_{th}=4m_q^2$.

The temperature and density effects on the total cross section for three values of energy $s$ are given in Fig.4. At low density, the chiral phase transition is washed out in the case of a non-vanishing current quark mass $m_0 = 0.005 GeV$, and so is the singularity in the cross section at the transition point. Even an expected maximum due to the small quark mass at this point does not exist. It reflects the chiral phenomenon that when the quark mass drops down with the increase of the temperature, the pion mass goes up simultaneously, and therefore it becomes more and more difficult for the quark plasma to hadronize into pions. At low temperature, while the first-order phase-transition remains, the dynamical singularity in the cross section is still washed out since the finite quark and pion masses. However, very different from the temperature dependence shown in Fig.4a where the quark plasma hadronizes smoothly and reaches the maximum rate within a temperature range $\Delta T\sim 0.1\ GeV$, there is a cliffy boundary in the $T-n_b$ plane (Fig.4b) separating the quark-meson plasma at low density from the quark plasma at high density. This feature arises from the discontinuities of the quark and pion masses at the first-order phase-transition point. When the density decreases, the quark plasma first meets the hadronization point and begins to hadronize, and then it reaches the first-order phase transition, the sudden increase of the quark mass combined with the sudden decrease of the pion mass at the transition point leads to a jump in the cross section. 

At zero temperature, the hadronization point and the chiral transition point are very close to each other, see in Fig.1, the quark plasma reaches the transition point immediately after the beginning of hadronization. Therefore, we see in Fig.4b a precipitous jump from zero to a maximum for any $s$. The maximum, although it is not strong, is also a consequence of the discontinuities of the quark and pion masses at the transition point.

\section { Conclusions }

The present investigation of the hadronization of a quark plasma in the NJL model to the first order in $1/N_c$ has been motivated by two aspects: 1) The chiral symmetry restoration in the chiral limit and its consequence in hadronization rate, a dynamical singularity at the critical point, are fully washed out when a non-vanishing current quark mass is taken into account. An important question is then what the influence of chiral dynamics is in the hadronization in the real quark world. 2) While the chiral phase transition happens at both high temperature and at high density, the order of the phase transition is different in two cases. The second-order phase transition due to temperature effect becomes a smooth crossover, but the first-order transition due to density effect is retained when the explicit chiral symmetry is considered. If such a first-order transition occurs in relativistic heavy ion collisions, we need to know what its experimental signatures are. 

The NJL model may be used as a model of QCD as long as one is interested in the chiral properties of the quark world. In this model the hadronization point of the interacting quarks is above the chiral phase transition point in the real world at high density, and the two points approach each other when the current quark mass goes to zero. Above the hadronization point one has a deconfined phase of quarks, but a quark-meson plasma below the point. Therefore, in the process of hadronization of the quark plasma, we expect that the retained first-order chiral transition at high density may still play an important role in the hadronization rate. 

We have calculated the energy, temperature and density dependence of the total hadronization cross section of a single quark in the framework of flavor $SU(2)$ NJL model. Corresponding to the motivations mentioned above, our main conclusions are: 1) The hadronization cross section is normally strongest for small values of the energy of the $q\bar q$ pair, and the singularity structure at low energies depends on the temperature and density. At low temperature or density, there are two singularities at the threshold energy and at the pole position ($m_\sigma^2$) of the $\sigma$ propagator. As the temperature or density increases, the kinematical singularity at the threshold energy vanishes first when the pion mass $m_\pi$ exceeds the quark mass $m_q$, and then the singularity at the pole disappears too when the threshold energy of the hadronization process, $s=4m_\pi^2$, is larger than $m_\sigma^2$. As for the critical scattering, the divergence at the critical point for all values of $s$ in the chiral limit is fully washed away. 2) Very different from the temperature effect which leads to a gradual hadronization during the cooling process of the quark plasma, the hadronization cross section is fairly constant in the process of hadronization at low temperature. The sudden jump of the cross section at the beginning of the hadronization means that the first-order transition of chiral symmetry restoration at high density is still reflected in the hadronization in the real quark world. A direct consequence of this feature is a suddenly strong increase of pion number in the expansion process of relativistic heavy ion collisions, if chiral symmetry is restored in the collisions. Our qualitative conclusions here derived from the NJL model may be more general because the first-order transition of chiral symmetry restoration looks like an essential feature of a strong interacting system at high density.          

{\bf acknowledgments}:
We are grateful to S.P.Klevansky for careful reading of the manuscript. The work was supported in part by the 
NSFC under grant numbers 19845001 and 19925519, and by the Major State Basic Research Development Program under contract number G2000077407.

\newpage
{\bf Figure Captions}\\ \\ 
{\bf Fig.1}: 
The phase-transition line and hadronization line in $T-\mu$ plane (Fig.1a) and $T-n_b$ plane (Fig.1b). The dashed lines indicate a second-order transition while the thin solid lines indicate one of first order in the chiral limit. The point $P$ is tricritical. The thick solid lines with critical end-point $E$ are the lines of first-order transition with non-vanishing current quark mass. For the first-order transition with and without current quark mass, the mixed region is divided into regions of instability ($US$) and metastability ($MS$). The hadronization line separating the region of quark plasma from that of quark-meson plasma is indicated by thin solid lines in the two planes.\\

{\bf Fig.2}: 
The Feynman diagrams for the hadronization of a $q\bar q$ pair into two pions to the first order in $1/N_c$ expansion. The solid lines denote quarks and antiquarks, the dashed lines denote mesons.\\

{\bf Fig.3}:
The total hadronization cross section as a function of the collision energy $s$, for three different baryon densities (Fig.3a) at $T=0$ and for three different temperatures (Fig.3b) at $n_b = 0$.\\

{\bf Fig.4}:
The total hadronization cross section as a function of the temperature (Fig.4a) at $n_b = 0$ and as a function of the baryon density (Fig.4b) at $T=0$. In each case, the collision energy takes three different values. 
 
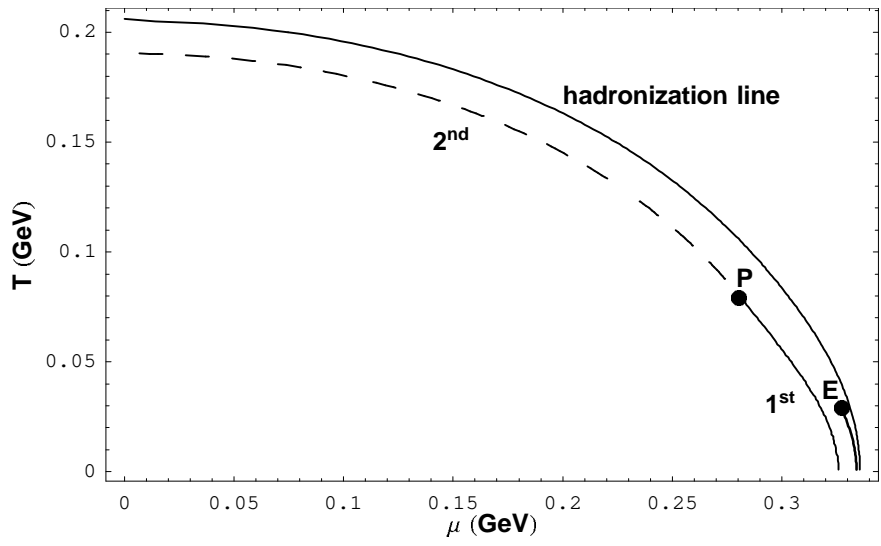
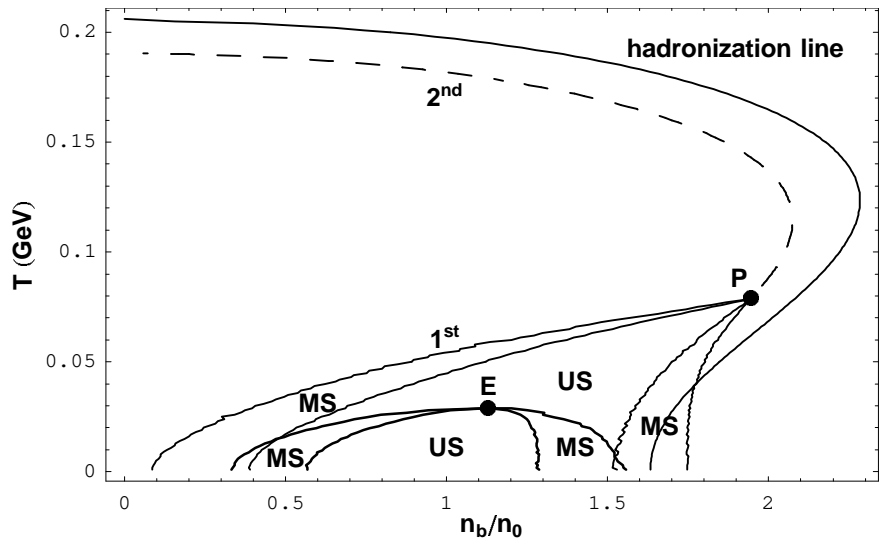
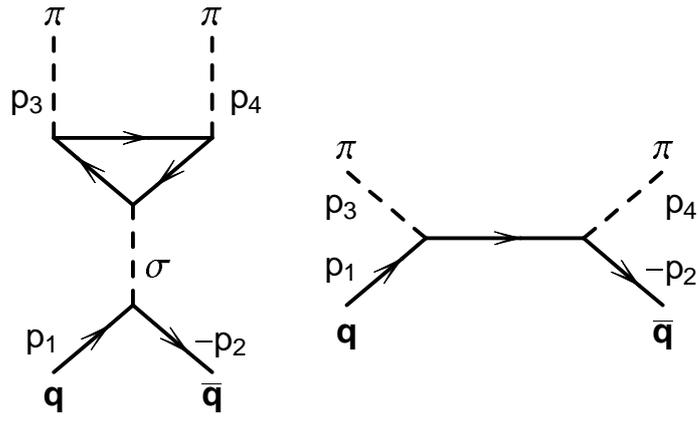
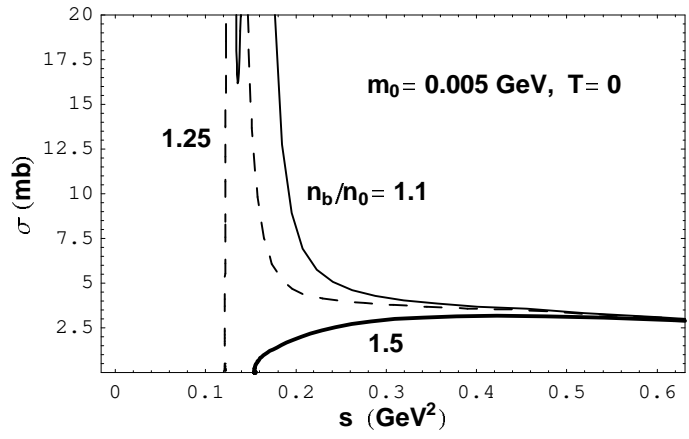
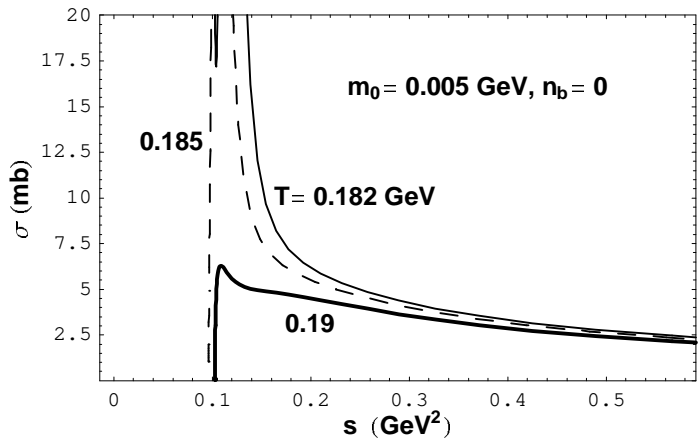
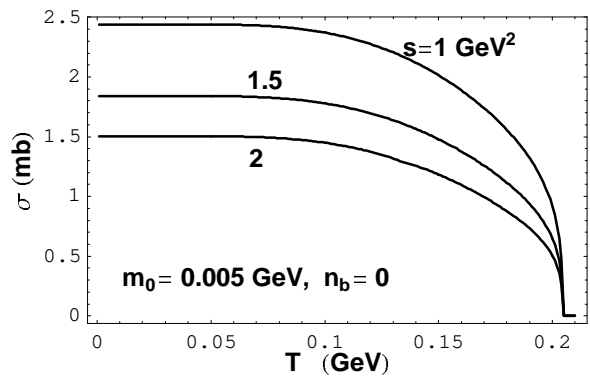
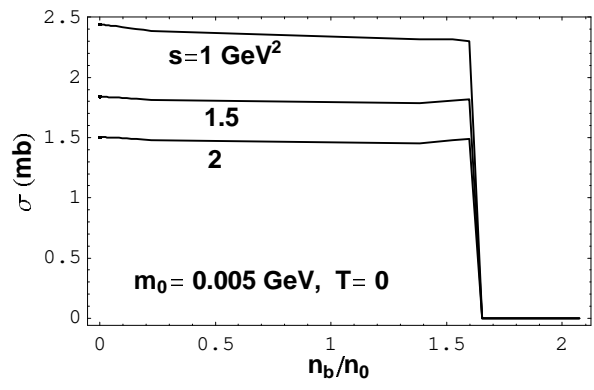
\end{document}